\documentclass[english]{article}
\usepackage{mathpazo}
\usepackage[T1]{fontenc}
\usepackage[latin9]{inputenc}
\usepackage{geometry}
\usepackage{color}
\usepackage{babel}
\usepackage{array}
\usepackage{amsmath}
\usepackage{amssymb}
\usepackage{graphicx}
\usepackage[unicode=true,
 bookmarks=true,bookmarksnumbered=false,bookmarksopen=false,
 breaklinks=false,pdfborder={0 0 1},backref=false,colorlinks=true]
 {hyperref}
\hypersetup{pdftitle={Contact preserving Riemann solver for incompressible two-phase flows-II},
 pdfauthor={Sourabh Bhat},
 pdfsubject={Contact preserving Riemann solver for incompressible two-phase flows},
 pdfkeywords={Riemann solver, Incompressible flow, VOF, HLLC-type}}

\makeatletter

\providecommand{\tabularnewline}{\\}
\newcommand{\lyxdot}{.}

\usepackage{cite}
\renewcommand{\fnum@figure}{Fig.~\thefigure}
\usepackage[font={footnotesize, bf}]{caption}

\makeatother

\begin{document}

\title{Modified HLLC-VOF solver for incompressible two-phase fluid flows}

\maketitle
\begin{center}
\begin{tabular}{>{\centering}p{0.5\textwidth}>{\centering}p{0.5\textwidth}}
Sourabh P. Bhat & J. C. Mandal\tabularnewline
\texttt{spbhat.mail@gmail.com}

\texttt{\href{https://spbhat.in/}{https://spbhat.in/}} & \texttt{mandal@aero.iitb.ac.in}

\texttt{https://www.aero.iitb.ac.in/\textasciitilde{}mandal/}\tabularnewline
\end{tabular}
\par\end{center}

\begin{center}
Department of Aerospace Engineering,\\
Indian Institute of Technology Bombay, Mumbai - 400076.
\par\end{center}
\begin{abstract}
A modified HLLC-type contact preserving Riemann solver for incompressible
two-phase flows using the artificial compressibility formulation is
presented. Here, the density is omitted from the pressure evolution
equation. Also, while calculating the eigenvalues and eigenvectors,
the variations of the volume fraction is taken into account. Hence,
the equations for the intermediate states and the intermediate wave
speed are different from the previous HLLC-VOF formulation {[}Bhat
S P and Mandal J C, \emph{J. Comput. Phys.} 379 (2019), pp. 173-191{]}.
Additionally, an interface compression algorithm is used in tandem
to ensure sharp interfaces. The modified Riemann solver is found to
be robust compared to the previous HLLC-VOF solver, and the results
produced are superior compared to non-contact preserving solver. Several
test problems in two- and three-dimensions are solved to evaluate
the efficacy of the solver on structured and unstructured meshes.
\end{abstract}

\section{Introduction}

The artificial compressibility formulation is a commonly accepted
way of solving single-phase incompressible fluid flows \cite{Belov1995,Tamamidis1996,Drikakis1998,Gaitonde1998,Manzari1999,Zhao2000,Malan2002,Mandal2009}.
However, the artificial compressibility formulation is less common
in solving multiphase incompressible flows \cite{Zhao2002,Nourgaliev2004,Yakovenko2008,Niu2011}.
In the original paper on artificial compressibility formulation by
Chorin \cite{Chorin1967}, the mass equation is converted into a pressure
evolution equation by adding an artificial term. This ingenious technique
allowed for the pressure to evolve with the velocity field in a closely-coupled
manner. The said paper and the others following this formulation use
a non-dimensional form of governing equations and hence do not include
density in the mass equation. However, when applying this technique
for solving two-phase flows there is no unique density to normalize,
and hence there are two possibilities. One is to use the conservative
variable in mass equation as $p/\left(\rho\beta\right)$ \cite{Niu2011,Bhat2019}
and the other method is to omit the density and use $p/\beta$ \cite{Zhao2002,Price2006,Yang2014,Parameswaran2019},
where $p$ is the pressure, $\rho$ is the fluid density and $\beta$
is the artificial compressibility parameter. In our recent paper \cite{Bhat2019},
we presented a contact preserving HLLC-type Riemann solver, for incompressible
two-phase flows using $p/\left(\rho\beta\right)$ in the mass equation,
which was called HLLC-VOF solver. In this follow-up paper, a contact
preserving Riemann solver for the second type of system using $p/\beta$
is presented. This modification seems to result in a more robust solver.
Because, in contrast to HLLC-VOF solver, the modified solver works
well with a single chosen value of $\beta$, in all the considered
test problems. Also, in the modified Riemann solver, the density in
the momentum variables is written as a function of volume fraction
when evaluating the eigenvalues and eigenvectors of the flux Jacobian.
Hence, the convective flux obtained by the modified Riemann solver
is considerably different compared to the previous HLLC-VOF solver.

Calculation of the convective flux using a Riemann solver (instead
of a geometric method such as the PLIC \cite{Rider1998}) has many
advantages, as Riemann solvers are independent of cell geometry. The
formulation remains same irrespective of the cell type, geometry complexity
and the number of physical dimensions. However, due to the upwind
nature of Riemann solvers, some numerical dissipation inherently gets
added to the flux, which helps in stabilizing the numerical scheme.
The HLLC-type contact-preserving Riemann solvers are less dissipative
and do not smear the fluid interface much, for small duration of evolution.
Therefore, in the earlier paper \cite{Bhat2019} this issue was not
addressed. However, this is not acceptable in problems evolving for
a long duration, or where the fluid interfaces move very slowly. This
is a well-known shortcoming and few techniques are available in literature
to mitigate this problem, such as the FCT method \cite{Rudman1997}
and the CICSAM method \cite{Ubbink1999}. Another approach for reducing
the numerical dissipation is to superimpose a compressive velocity
field over the existing velocity field, such that the interface thickness
gets restricted. This is done by adding an artificial compressive
term to the VOF equation \cite{Rusche2002,Cifani2016}, which is active
only near the interface. This method is used in OpenFOAM software
\cite{Jasak2009} and further advances are made to this method to
ensure correct compressive flux \cite{Lee2015,Mehmani2018}. In this
paper, the interface compression methodology is adopted as it fits
naturally within the underlying framework.

To evaluate the efficacy of the modified solver, which we refer to
as HLLC-VOF-M, various problems are solved on structured and unstructured
mesh. The problems are chosen systematically such that only a few
aspects of the solver are tested at a time. The solutions produced
by the HLLC-VOF-M solver are compared with a non-contact capturing
solver, experimental, theoretical and numerical solutions depending
on availability of results in the literature. A three-dimensional
problem involving merging of complex interfaces is also solved to
demonstrate the extensibility of the present solver.

The remaining paper is organized as follows. We begin by describing
the governing equations for two-phase incompressible fluid flows in
the next section. The numerical formulation is given in section~\ref{sec:Numerical-formulation},
and it comprises of nine sub-sections: finite volume method, treatment
of time derivatives, convective flux, interface compression, viscous
flux, surface tension, source term, initial conditions, and boundary
conditions. Out of these nine sub-sections, our main contribution
is in sub-section~\ref{sub:Riemann-solver}, where a detailed description
of the HLLC-VOF-M Riemann solver is presented for calculation of the
convective flux. Section~\ref{sec:The-Algorithm} provides a brief
overview and a flowchart of the complete algorithm used by the solver.
Five numerical test problems, of varying nature, are presented and
discussed in section~\ref{sec:Results-and-discussion}. Finally,
we conclude in section~\ref{sec:Conclusion}, by summarizing the
key contributions from the paper.

\section{\label{sec:Governing-equations}Governing equations}

The governing equations for 3-dimensional, unsteady incompressible
two-phase flows, using dual-time stepping, artificial compressibility
formulation and volume of fluid method, can be written as, 
\begin{equation}
\begin{array}{c}
{\displaystyle \frac{\partial\mathbf{U}}{\partial\tau}+\frac{\partial\mathbf{W}}{\partial t}+\frac{\partial\mathbf{F}}{\partial x}+\frac{\partial\mathbf{G}}{\partial y}+\frac{\partial\mathbf{H}}{\partial z}+\nabla\cdot\mathbf{F}_{c}=\frac{\partial\mathbf{F_{v}}}{\partial x}+\frac{\partial\mathbf{G_{v}}}{\partial y}+\frac{\partial\mathbf{H_{v}}}{\partial z}+\nabla\cdot\mathbf{T}+\mathbf{S}};\\
\mathbf{U}=\left[\begin{array}{c}
p/\beta\\
\rho u\\
\rho v\\
\rho w\\
C
\end{array}\right],\ \mathbf{W}=\left[\begin{array}{c}
0\\
\rho u\\
\rho v\\
\rho w\\
C
\end{array}\right],\ \mathbf{F}=\left[\begin{array}{c}
u\\
\rho u^{2}+p\\
\rho uv\\
\rho uw\\
uC
\end{array}\right],\ \mathbf{G}=\left[\begin{array}{c}
v\\
\rho uv\\
\rho v^{2}+p\\
\rho vw\\
vC
\end{array}\right],\ \mathbf{H}=\left[\begin{array}{c}
w\\
\rho uw\\
\rho vw\\
\rho w^{2}+p\\
wC
\end{array}\right],\\
\mathbf{F_{v}}=\left[\begin{array}{c}
0\\
2\mu\frac{\partial u}{\partial x}\\
\mu\left(\frac{\partial u}{\partial y}+\frac{\partial v}{\partial x}\right)\\
\mu\left(\frac{\partial u}{\partial z}+\frac{\partial w}{\partial x}\right)\\
0
\end{array}\right],\ \mathbf{G_{v}}=\left[\begin{array}{c}
0\\
\mu\left(\frac{\partial u}{\partial y}+\frac{\partial v}{\partial x}\right)\\
2\mu\frac{\partial v}{\partial y}\\
\mu\left(\frac{\partial v}{\partial z}+\frac{\partial w}{\partial y}\right)\\
0
\end{array}\right],\ \mathbf{H_{v}}=\left[\begin{array}{c}
0\\
\mu\left(\frac{\partial u}{\partial z}+\frac{\partial w}{\partial x}\right)\\
\mu\left(\frac{\partial v}{\partial z}+\frac{\partial w}{\partial y}\right)\\
2\mu\frac{\partial w}{\partial z}\\
0
\end{array}\right],\ \mathbf{S}=\left[\begin{array}{c}
0\\
\rho g_{x}\\
\rho g_{y}\\
\rho g_{z}\\
0
\end{array}\right];
\end{array}\label{eq:AC_VOF-NS}
\end{equation}
where, $\tau$ is a pseudo-time variable used for iterating to a converged
solution at a particular real-time $t$. $\mathbf{U}$ is the conservative
variable vector updated in pseudo-time, $\mathbf{W}$ is the conservative
variable vector updated in real-time, $\left(\mathbf{F},\mathbf{G},\mathbf{H}\right)$
are convective flux vectors, $\nabla\cdot\mathbf{F}_{c}$ is the interface
compression term, $\left(\mathbf{F_{v}},\mathbf{G_{v}},\mathbf{H_{v}}\right)$
are the viscous flux vectors, $\mathbf{S}$ is a vector containing
source terms and $\mathbf{T}$ is a tensor for surface tension force.
The variable $p$ denotes pressure, $\beta$ is the artificial compressibility
parameter, $\left(u,v,w\right)$ is the velocity vector in $\left(x,y,z\right)$
Cartesian space co-ordinates, $\rho$ is the density of the fluid,
$\mu$ is the dynamic viscosity of the fluid and $\left(g_{x},g_{y},g_{z}\right)$
is the acceleration due to gravity. It may be noted that the conservative
variable $p/\beta$ in the mass equation is an artificial term added
for coupling the mass and momentum equations. Also, this modification
converts the pseudo-time derivative along with the convective flux
into a hyperbolic system. Hence, Riemann solvers can be used for calculation
of convective flux.

The numerical simulation of two-phase flow, brings in an additional
fifth equation for advection of volume fraction, $C$, and a tensor
associated with surface tension $\mathbf{T}$. The details of surface
tension are explained in sub-section~\ref{sub:surface-tension}.
An interface compression term, $\nabla\cdot\mathbf{F}_{c}$ is introduced
to ensure sharp interfaces, the details of which are given in sub-section
\ref{sub:interface-compression}. The density, $\rho$, and the dynamic
viscosity, $\mu$, are defined as a function of volume fraction as,
\begin{equation}
\rho=\rho\left(C\right)=\rho_{1}\, C+\rho_{2}\,\left(1-C\right)=\left(\rho_{1}-\rho_{2}\right)\, C+\rho_{2},\label{eq:density_function}
\end{equation}
\begin{equation}
\mu=\mu\left(C\right)=\mu_{1}\, C+\mu_{2}\,\left(1-C\right)=\left(\mu_{1}-\mu_{2}\right)\, C+\mu_{2}\label{eq:dyn_visc_function}
\end{equation}
where, $\rho_{1}$ is the density of first fluid, $\rho_{2}$ is the
density of the second fluid, $\mu_{1}$ is the dynamic viscosity of
the first fluid and $\mu_{2}$ is the dynamic viscosity of the second
fluid.

\section{\label{sec:Numerical-formulation}Numerical formulation}

\subsection{Finite volume method}

\begin{figure}[h]
\noindent \begin{centering}
\includegraphics[width=0.7\textwidth,draft=false]{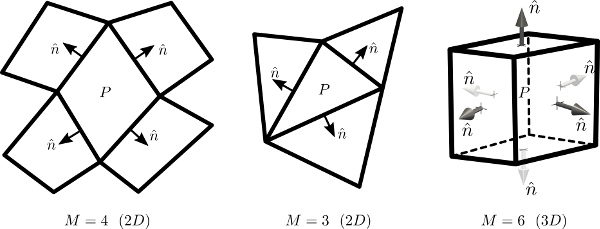}
\par\end{centering}

\caption{\label{fig:fv-cells}Schematic diagrams for displaying the notations
used in finite volume formulation.}

\end{figure}

The finite volume discretization of the governing equations (\ref{eq:AC_VOF-NS})
is obtained by integrating the equations over an arbitrary volume
in space. The volume of a cell $P$ is denoted as $\Omega_{P}$. Different
types of cells used in this work are shown in Fig.~\ref{fig:fv-cells}.
Using the Gauss-divergence theorem, the volume integrals reduce to
surface integrals resulting in the following discretized form of the
governing equations.

\noindent 
\begin{multline}
\Omega_{P}\,\frac{\partial\overline{\mathbf{U}}}{\partial\tau}+\Omega_{P}\,\frac{\partial\overline{\mathbf{W}}}{\partial t}+\sum_{m=1}^{M}\left(\mathbf{F}\, n_{x}+\mathbf{G}\, n_{y}+\mathbf{H}\, n_{z}\right)_{m}\,\Gamma_{m}+\sum_{m=1}^{M}\left(\mathbf{F}_{c}\cdot\hat{n}\,\Gamma\right)_{m}\\
=\sum_{m=1}^{M}\left(\mathbf{F_{v}}\, n_{x}+\mathbf{G_{v}}\, n_{y}+\mathbf{H_{v}}\, n_{z}\right)_{m}\,\Gamma_{m}+\sum_{m=1}^{M}\left(\mathbf{T}\cdot\hat{n}\,\Gamma\right)_{m}+\Omega_{P}\,\overline{\mathbf{S}}\ .\label{eq:fv-formulation}
\end{multline}
The over-bar indicates cell averaged value of a variable inside cell
$P$. The cell is enclosed by $M$ planar faces, denoted by $m=1\dots M$,
with the $m^{\text{th}}$ face having a surface area of $\Gamma_{m}$.
The unit normal vector $\hat{n}=\left(n_{x},n_{y},n_{z}\right)$ for
each of the planar faces points outward of the cell $P$. It may be
noted that, in case of a second order accurate finite volume method,
the cell-averaged values can be obtained by evaluating the variables
at cell-centroid \cite{V.Venkatakrishnan1995}.

\noindent 

\subsection{Treatment of pseudo and real time derivatives}

The dual-time stepping procedure proposed by Jameson \cite{Jameson1991}
for compressible flows can be utilized along with the artificial compressibility
formulation to obtain a time accurate solution for incompressible
flows. The basic pseudo-compressibility procedure followed here is
similar to the one described by Gaitonde \cite{Gaitonde1998}. The
finite volume formulation (\ref{eq:fv-formulation}) can be re-written
as an ordinary differential equation (ODE) in pseudo-time as, 
\begin{equation}
\Omega_{P}\frac{d\overline{\mathbf{U}}}{d\tau}=-R\left(\overline{\mathbf{U}}\right)\ ,\label{eq:fv-formulation-1}
\end{equation}
where, $R\left(\overline{\mathbf{U}}\right)$ is called the residual
of cell $P$ and can be written as,
\begin{multline}
R\left(\overline{\mathbf{U}}\right)=\Omega_{P}\,\frac{\partial\overline{\mathbf{W}}}{\partial t}+\sum_{m=1}^{M}\left(\mathbf{F}\, n_{x}+\mathbf{G}\, n_{y}+\mathbf{H}\, n_{z}\right)_{m}\,\Gamma_{m}+\sum_{m=1}^{M}\left(\mathbf{F}_{c}\cdot\hat{n}\,\Gamma\right)_{m}\\
-\sum_{m=1}^{M}\left(\mathbf{F_{v}}\, n_{x}+\mathbf{G_{v}}\, n_{y}+\mathbf{H_{v}}\, n_{z}\right)_{m}\,\Gamma_{m}-\sum_{m=1}^{M}\left(\mathbf{T}\cdot\hat{n}\,\Gamma\right)_{m}-\Omega_{P}\,\overline{\mathbf{S}}\ .
\end{multline}
Based on the ODE given by equation (\ref{eq:fv-formulation-1}), the
solution is updated in pseudo-time using the two-stage stability preserving
Runge-Kutta method \cite{Gottlieb2009}, which can be written as,\renewcommand{\arraystretch}{1.5}
\begin{equation}
\begin{array}{rcl}
\mathbf{U}^{\left(0\right)} & = & \mathbf{\overline{U}}^{k};\\
\mathbf{U}^{\left(1\right)} & = & \mathbf{U}^{\left(0\right)}-\frac{\Delta\tau}{\Omega_{P}}\, R\left(\mathbf{U}^{\left(0\right)}\right);\\
\mathbf{U}^{\left(2\right)} & = & \frac{1}{2}\,\mathbf{U}^{\left(0\right)}+\frac{1}{2}\,\mathbf{U}^{\left(1\right)}-\frac{1}{2}\,\frac{\Delta\tau}{\Omega_{P}}\, R\left(\mathbf{U}^{\left(1\right)}\right);\\
\mathbf{\overline{U}}^{k+1} & = & \mathbf{U}^{\left(2\right)};
\end{array}\label{eq:RK-3}
\end{equation}
\renewcommand{\arraystretch}{1}where, $\mathbf{U}^{\left(1\right)}$,
$\mathbf{U}^{\left(2\right)}$ are the solutions at intermediate pseudo-time
steps, $\mathbf{\overline{U}}^{k}$ is the solution at current pseudo-time
and $\mathbf{\overline{U}}^{k+1}$ is the updated solution in pseudo-time.
The pseudo-time iterations are carried out until the $L_{2}$-norm
of the residual is reduced to a very small value, of $10^{-3}$, for
all the conservative variables. Thus recovering the original unsteady
two-phase flow equations, such that $\partial\overline{\mathbf{U}}/\partial\tau\approx0$.
The $L_{2}$-norm of the residual is defined as given by Wang et.~al
\cite{Wang2013}. It may be noted that all the above variable vectors
are functions of $\overline{\mathbf{U}}$, i.e. $\overline{\mathbf{W}},\mathbf{F},\mathbf{G},\mathbf{H},\mathbf{F}_{c},\mathbf{F_{v}},\mathbf{G_{v}},\mathbf{H_{v}},\mathbf{T}$
and $\overline{\mathbf{S}}$ are dependent on $\overline{\mathbf{U}}$.
Hence, updating $\overline{\mathbf{U}}$ will modify the residual
at every Runge-Kutta step.

The real-time derivative is numerically approximated using a three-point
implicit backward difference formula, which can be easily derived
to be,
\begin{equation}
\frac{\partial\overline{\mathbf{W}}}{\partial t}=\frac{3\,\overline{\mathbf{W}}^{k}-4\,\overline{\mathbf{W}}^{n}+\overline{\mathbf{W}}^{n-1}}{2\Delta t}\ .
\end{equation}
where, $\Delta t$ is the chosen time step to evolve the solution
in real-time (this time step is problem dependent and is defined for
each problem in section \ref{sec:Results-and-discussion}, namely
Results and discussion), $k$ denotes the current solution in pseudo-time,
$n$ denotes the current solution in real-time, $n-1$ denotes the
previous solution in real-time. The solution at previous real-time
level is not available at the beginning of the simulation, hence,
a two-point backward difference approximation is used. The two-point
backward difference approximation can be written as,
\begin{equation}
\frac{\partial\overline{\mathbf{W}}}{\partial t}=\frac{\overline{\mathbf{W}}^{k}-\overline{\mathbf{W}}^{n}}{\Delta t}\ .
\end{equation}

The pseudo-time step size, $\Delta\tau$, is limited by the stability
requirements of the scheme. The pseudo-time step is dependent on the
convective flux, viscous flux and the surface tension flux. Therefore,
\begin{equation}
\Delta\tau\leq\min\left(\Delta\tau_{cv},\Delta\tau_{s}\right)
\end{equation}
where, $\Delta\tau_{cv}$ is the time step constraint due to convective
and viscous flux, and $\Delta\tau_{s}$ is the time step constraint
imposed by surface tension. The largest time step for a cell $P$
due to convective and viscous fluxes may be estimated using \cite{Mavriplis1990,Swanson1991,Blazek2015-c6}
\begin{equation}
\left(\Delta\tau_{cv}\right)_{P}=\frac{\Omega_{P}}{\left(\Lambda_{c}+K\Lambda_{v}\right)_{P}}\ ,
\end{equation}
where, $\Lambda_{c}$ and $\Lambda_{v}$ are estimates of the convective
and viscous spectral radii for the cell respectively. The spectral
radii are calculated as,
\begin{equation}
\Lambda_{c}=\sum_{m=1}^{M}\left(\max\left|\lambda\right|\,\Gamma\right)_{m}\quad\text{ and }\quad\Lambda_{v}=\frac{4}{3\,\Omega_{P}}\sum_{m=1}^{M}\left(\frac{\mu}{\rho}\Gamma^{2}\right)_{m}\ ,
\end{equation}
where, $\max\left|\lambda\right|$ is the maximum absolute eigenvalue
at the face $m$, $K=4$ is an empirically determined coefficient
in \cite{Mavriplis1990}. The time constraint due to surface tension
is estimated based on the expressions given in \cite{Brackbill1992,Tryggvason2011}
to be,
\begin{equation}
\Delta\tau_{s}=\left[\frac{\left(\rho_{1}+\rho_{2}\right)\Omega_{P}}{\pi\,\sigma}\right]^{1/2}\ .
\end{equation}
The pseudo-time step is further limited by the real-time step, $\Delta t$,
which is estimated using linear stability analysis in \cite{Arnone1993}
as,
\begin{equation}
\Delta\tau\leq\frac{2}{3}\Delta t\ .
\end{equation}
Finally a scaling factor, similar to Courant number, is used to obtain
the following expression for pseudo-time step,
\begin{equation}
\Delta\tau=\mbox{CFL}\left[\min\left(\Delta\tau_{cv},\Delta\tau_{s},\frac{2}{3}\Delta t\right)\right]\ .
\end{equation}
The scaling factor of $\text{CFL}=1$ is used in all the simulations
in this paper. A local-time stepping algorithm is used to accelerate
the convergence, where the value of $\Delta\tau$ may be different
for each cell based on the local flow field.

\subsection{Calculation of convective flux}

The novelty of this paper lies in the development of an HLLC-type
Riemann solver, for calculation of convective flux for three-dimensional,
incompressible two-phase flows, governed by equations (\ref{eq:AC_VOF-NS}).
The finite volume formulation provides a discrete, cell-averaged conservative
variable vector, $\overline{\mathbf{U}}$, for each cell. Using these
discrete solution vectors, the calculation of the convective flux
is done in two steps.
\begin{enumerate}
\item \emph{Solution reconstruction}, where the discrete solution vectors
in the neighborhood of the cell, are used to obtain a smooth distribution
of solution vector inside each cell. This will potentially result
in discontinuities in solution vectors at the cell faces, leading
to Riemann problems at cell faces.
\item Solving the \emph{Riemann problem} to compute the convective flux,
using the newly developed HLLC-VOF-M Riemann solver, at the cell faces.
\end{enumerate}
These two steps are explained in the next two sub-sections.

\subsubsection{Solution reconstruction}

\begin{figure}[h]
\begin{centering}
\includegraphics[width=0.5\textwidth,draft=false]{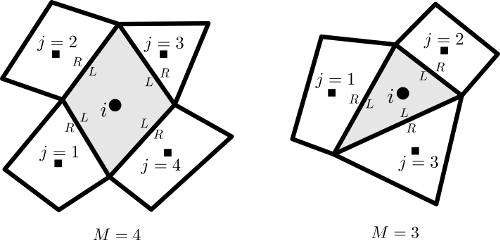}
\par\end{centering}

\caption{\label{fig:mixed-cells}Subset of computational grids consisting of
mixed cell-types. The centroid of the cell in which the solution is
being reconstructed is represented using {\Large{}$\bullet$} symbol
and centroid of neighboring cells is represented using {\footnotesize{}$\blacksquare$}
symbol.}
\end{figure}

The conservative variables are reconstructed, using the local cell
averaged data, to obtain a smooth distribution inside each cell. This
is done by using a truncated Taylor series approximation and a distance
weighted least-square approach. Consider a subset of a computational
grid, as shown in Fig.~\ref{fig:mixed-cells}.

\noindent 

\noindent The cell, in which the solution reconstruction is carried
out, is denoted as $i$ and the face-based neighboring cells are denoted
as $j=1,2\dots M$. Let a scalar $\xi$ represent a single component
from vector $\mathbf{U}$. A truncated Taylor series, about the centroid
of cell $i$, can be written for approximating $\xi$ at the centroid
of cell $j$ as,
\begin{equation}
\xi_{j}=\xi_{i}+\left.\frac{\partial\xi}{\partial x}\right|_{i}\Delta x_{j}+\left.\frac{\partial\xi}{\partial y}\right|_{i}\Delta y_{j}+\left.\frac{\partial\xi}{\partial z}\right|_{i}\Delta z_{j}
\end{equation}
or the change in $\xi$ can be written as, 
\begin{equation}
\Delta\xi_{j}=\left.\frac{\partial\xi}{\partial x}\right|_{i}\Delta x_{j}+\left.\frac{\partial\xi}{\partial y}\right|_{i}\Delta y_{j}+\left.\frac{\partial\xi}{\partial z}\right|_{i}\Delta z_{j}
\end{equation}
where, $\Delta\xi_{j}=\left(\xi_{j}-\xi_{i}\right)$, $\Delta x_{j}=\left(x_{j}-x_{i}\right)$,
$\Delta y_{j}=\left(y_{j}-y_{i}\right)$, $\Delta z_{j}=\left(z_{j}-z_{i}\right)$.
The derivatives of $\xi$ are evaluated at the centroid of cell $i$.
The second and higher order terms of the Taylor series are neglected,
which results in a linear approximation. Writing the equations for
$j=1\dots M$, produces $M$ linear equations and may be written as
a over-determined system of equations,
\begin{equation}
\underbrace{\left[\begin{array}{ccc}
\Delta x_{1} & \Delta y_{1} & \Delta z_{1}\\
\Delta x_{2} & \Delta y_{2} & \Delta z_{2}\\
\vdots & \vdots & \vdots\\
\Delta x_{M} & \Delta y_{M} & \Delta z_{M}
\end{array}\right]}_{S}\underbrace{\left[\begin{array}{c}
\left(\partial\xi/\partial x\right)_{i}\\
\left(\partial\xi/\partial y\right)_{i}\\
\left(\partial\xi/\partial z\right)_{i}
\end{array}\right]}_{\nabla\xi_{i}}=\underbrace{\left[\begin{array}{c}
\Delta\xi_{1}\\
\Delta\xi_{2}\\
\vdots\\
\Delta\xi_{M}
\end{array}\right]}_{\Delta\xi}\ .
\end{equation}
An inverse-distance based weights, $w_{j}=1/\left\Vert \Delta r_{j}\right\Vert _{2}$,
are used to increase the influence of the nearby neighbors, where
$\left\Vert \Delta r_{j}\right\Vert _{2}$ is the Euclidean distance
between the points $i$ and $j$, calculated as,
\begin{equation}
\left\Vert \Delta r_{j}\right\Vert _{2}=\sqrt{\Delta x_{j}^{2}+\Delta y_{j}^{2}+\Delta z_{j}^{2}}\ .
\end{equation}
 The resulting weighted matrix may be written as,
\begin{equation}
W\, S\,\nabla\xi_{i}=W\,\Delta\xi\ ,
\end{equation}
where, $W=\text{diag}\left(w_{1},w_{2},\dots,w_{M}\right)$. This
over-determined system of equations is solved for $\nabla\xi_{i}$
using the singular value decomposition (SVD) method, using an open-source
software library \cite{ApacheCommonsMath}. The SVD method has an
advantage that its formulation remains same for 2D and 3D grids and
it can also deal with highly skewed grids very well. The method is,
however, computationally expensive compared to other direct methods.
This is of a less concern here, as the decomposition needs to be computed
only once for problems involving stationary grids. The computed derivatives
are modified using the Venkatakrishnan slope limiter \cite{V.Venkatakrishnan1995}
to avoid spurious oscillations near the interface, due to the discontinuous
nature of the volume fraction. The final linearly reconstructed variable,
$\xi$, inside the cell, may be written as,
\begin{equation}
\xi\left(x,y,z\right)=\xi_{i}+\mathbf{\Phi}_{i}\nabla\xi_{i}\cdot\Delta r\ ,
\end{equation}
where, $\left(x,y,z\right)$ is a point within or on the boundary
of cell $i$, $\mathbf{\Phi}_{i}$ is called the limiter for cell
$i$ and $\Delta r=\left(\Delta x,\Delta y,\Delta z\right)$ with
$\Delta x=\left(x-x_{i}\right)$, $\Delta y=\left(y-y_{i}\right)$,
$\Delta z=\left(z-z_{i}\right)$. The limiter is defined as \cite{V.Venkatakrishnan1995},
\begin{equation}
\mathbf{\Phi}_{i}=\min\left(\Phi_{1},\Phi_{2},\dots\Phi_{M}\right)\ ,
\end{equation}
where, for each of the face, 
\begin{equation}
\Phi_{j}=\begin{cases}
\phi\left(\frac{\xi_{i}^{\max}-\xi_{i}}{\xi_{j}-\xi_{i}}\right), & \text{if }\ \xi_{j}-\xi_{i}>0\\
\phi\left(\frac{\xi_{i}^{\min}-\xi_{i}}{\xi_{j}-\xi_{i}}\right), & \text{if }\ \xi_{j}-\xi_{i}<0\\
1, & \text{if }\ \xi_{j}-\xi_{i}=0\ .
\end{cases}
\end{equation}
The smooth function, $\phi\left(\eta\right)$, is given by \cite{V.Venkatakrishnan1995},
\begin{equation}
\phi\left(\eta\right)=\frac{\eta^{2}+2\eta}{\eta^{2}+\eta+2}\ .
\end{equation}
The reconstruction of solution data within each cell can potentially
create discontinuities at the cell faces, resulting in a Riemann problem
at each cell face. The reconstructed solution at a face, therefore
has two values, which are commonly called as the left state and the
right state. This is depicted in Fig.~\ref{fig:mixed-cells} using
symbols $L$ and $R$ respectively. The left and the right states
are computed at the centroid of the face to obtain a second-order
accurate solution reconstruction. Upon reconstruction of all the components
of vector $\mathbf{U}$, using the above procedure, the left and right
states, $\mathbf{U}_{L}$ and $\mathbf{U}_{R}$, are obtained at the
centroid of each face. The convective flux is calculated by using
a Riemann solver based on the obtained jump condition, as described
in the upcoming sub-section.

\subsubsection{\label{sub:Riemann-solver}Contact preserving Riemann solver}

In the present solver, a three-wave system is considered with the
left moving, right moving and intermediate wave. It is observed, with
the help of generalized Riemann invariant analysis that across the
intermediate wave there are jumps in the tangential component of velocity
and volume fraction, while the normal component of velocity and pressure
remains continuous. This is similar to the behavior displayed by the
contact wave in compressible fluid flows. It is well established in
case of compressible flows that Riemann solvers considering the intermediate
contact wave, such as HLLC \cite{Toro1994}, produce superior results.
Hence, a Riemann solver for incompressible two-phase flow, which is
derived considering the intermediate wave is expected to produce more
accurate results in flows involving fluid interfaces. A similar argument
is presented in the book by Toro \cite{Toro2009} for compressible
multi-component flows. Going ahead with this philosophy, a Riemann
solver is proposed for incompressible two-phase flows. The convective
flux in a finite volume formulation is computed using the following
three steps:
\begin{enumerate}
\item Given the reconstructed conservative variables for a face, $\left(\mathbf{U}_{L},\mathbf{U}_{R}\right)$,
the transformed variables, in a locally rotated coordinate system,
$\left(\mathbf{\hat{U}}_{L},\mathbf{\hat{U}}_{R}\right)=\left(T\,\mathbf{U}_{L},T\,\mathbf{U}_{R}\right)$
are calculated, which is elaborated below.
\item Using $\left(\mathbf{\hat{U}}_{L},\mathbf{\hat{U}}_{R}\right)$, the
Riemann flux, $\mathbf{\hat{F}}_{f}$, for each face of the cell is
calculated.
\item The convective flux required in the three-dimensional finite volume
formulation, $T^{-1}\mathbf{\hat{F}}_{f}$ is then calculated, which
can be added over the faces of the cell to obtain total flux for the
cell.
\end{enumerate}
The above steps are described in detail in the upcoming text.

\noindent 
\begin{figure}[h]
\begin{centering}
\includegraphics[width=0.15\textwidth,draft=false]{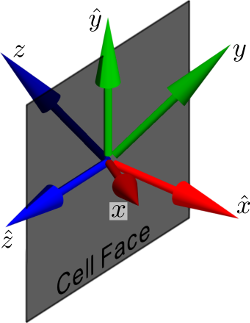}
\par\end{centering}

\caption{\label{fig:rotated-coordinate-system}Schematic diagram to display
the rotated coordinate system $\left(\hat{x},\hat{y},\hat{z}\right)$
in which the Riemann flux is computed.}
\end{figure}

\begin{figure}[h]
\begin{centering}
\includegraphics[width=0.4\textwidth,draft=false]{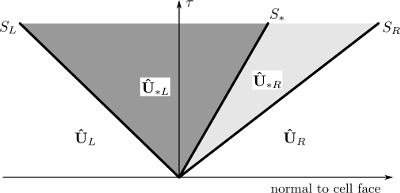}
\par\end{centering}

\caption{\label{fig:HLLC-VOF-wave-structure}Schematic diagram depicting the
wave structure of a Riemann solver with three distinct waves.}
\end{figure}

As the interest in this subsection is to obtain the convective flux,
let us consider a simplified system with only the pseudo-time derivative
and the convective flux vectors. Additionally, consider a rotated
coordinate system $\left(\hat{x},\hat{y},\hat{z}\right)$, as shown
schematically in Fig.~\ref{fig:rotated-coordinate-system}, where
$\hat{x}$ is the direction normal to the face and $\hat{y}$ and
$\hat{z}$ are tangential to the face. In such a coordinate system,
only the $\hat{x}$-split three-dimensional governing equations are
needed for calculation of convective flux, as the other two flux components
will not contribute in the finite volume flux computation. Thus, to
compute the convective flux using a Riemann solver the following simplified
system of governing equations is sufficient,
\begin{equation}
\frac{\partial\mathbf{\hat{U}}}{\partial\tau}+\frac{\partial\mathbf{\hat{F}}}{\partial\hat{x}}=0\ ,\label{eq:simplified-1d-eqn}
\end{equation}
where, $\mathbf{\hat{F}}=\mathbf{F}\left(\mathbf{\hat{U}}\right)$
and $\mathbf{\hat{U}}=T\,\mathbf{U}$, with the expression for $\mathbf{U}$
same as in equation (\ref{eq:AC_VOF-NS}). The transformation matrix
$T$ can be obtained to be, 
\begin{equation}
T=\left[\begin{array}{ccccc}
1 & 0 & 0 & 0 & 0\\
0 & n_{x} & n_{y} & n_{z} & 0\\
0 & t_{1x} & t_{1y} & t_{1z} & 0\\
0 & t_{2x} & t_{2y} & t_{2z} & 0\\
0 & 0 & 0 & 0 & 1
\end{array}\right]\ ,
\end{equation}
where, $\hat{n}=\left(n_{x},n_{y},n_{z}\right)$ is the unit vector
normal to the face, pointing from left state to right state (along
$\hat{x}$ direction), and, $\hat{t}_{1}=\left(t_{1x},t_{1y},t_{1z}\right)$
and $\hat{t}_{2}=\left(t_{2x},t_{2y},t_{2z}\right)$ are unit orthogonal
vectors tangent to the face. The transformed conservative variable
vector, $\mathbf{\hat{U}}$, can be written as,
\[
\mathbf{\hat{U}}=\left[\begin{array}{c}
p/\beta\\
\rho\,\left(u\, n_{x}+v\, n_{y}+w\, n_{z}\right)\\
\rho\,\left(u\, t_{1x}+v\, t_{1y}+w\, t_{1z}\right)\\
\rho\,\left(u\, t_{2x}+v\, t_{2y}+w\, t_{2z}\right)\\
C
\end{array}\right]=\left[\begin{array}{c}
p/\beta\\
\rho\,\hat{u}\\
\rho\,\hat{v}\\
\rho\,\hat{w}\\
C
\end{array}\right]\ ,
\]
where, $\hat{u}=u\, n_{x}+v\, n_{y}+w\, n_{z}$; $\hat{v}=u\, t_{1x}+v\, t_{1y}+w\, t_{1z}$
and $\hat{w}=u\, t_{2x}+v\, t_{2y}+w\, t_{2z}$. The Riemann solver
for calculation of convective flux takes the left and the right states
at the face, $\left(\mathbf{\hat{U}}_{L},\mathbf{\hat{U}}_{R}\right)$,
as inputs and returns the flux through the face, $\mathbf{\hat{F}}_{f}$
as an output.
\begin{equation}
\mathbf{\hat{F}}_{f}=\mathbf{F}_{f}\left(\mathbf{\hat{U}}_{L},\mathbf{\hat{U}}_{R}\right)\ ,
\end{equation}
where, $\mathbf{\hat{F}}_{f}$ is the Riemann flux at the cell face.
The two initial states, $\left(\mathbf{\hat{U}}_{L},\mathbf{\hat{U}}_{R}\right)$,
at the cell face, evolve over pseudo-time, $\tau$, into multiple
solution states spreading out in space and time, as shown in Fig.~\ref{fig:HLLC-VOF-wave-structure}.
These solution states are separated by discontinuities across the
waves. The speeds at which the waves move are associated with the
eigenvalues of the Jacobian, $\partial\mathbf{\hat{F}}/\partial\mathbf{\hat{U}}$,
which can be calculated to be, 
\begin{equation}
\frac{\partial\mathbf{\hat{F}}}{\partial\mathbf{\hat{U}}}=\left[\begin{array}{ccccc}
0 & 1/\rho & 0 & 0 & \left(\rho_{2}-\rho_{1}\right)\,\hat{u}/\rho\\
\beta & 2\,\hat{u} & 0 & 0 & \left(\rho_{2}-\rho_{1}\right)\,\hat{u}^{2}\\
0 & \hat{v} & \hat{u} & 0 & \left(\rho_{2}-\rho_{1}\right)\,\hat{u}\,\hat{v}\\
0 & \hat{w} & 0 & \hat{u} & \left(\rho_{2}-\rho_{1}\right)\,\hat{u}\,\hat{w}\\
0 & C/\rho & 0 & 0 & \rho_{2}\,\hat{u}/\rho
\end{array}\right]
\end{equation}
The eigenvalues of the Jacobian can be calculated to be,
\begin{equation}
\lambda=\left[\hat{u}_{C}-a,\hat{u},\hat{u},\hat{u},\hat{u}_{C}+a\right]
\end{equation}
where,
\begin{equation}
\hat{u}_{C}=\frac{\left(\rho+\rho_{2}\right)\,\hat{u}}{2\,\rho}=\frac{\hat{u}}{2}\left(1+\frac{\rho_{2}}{\rho}\right);\quad a=\sqrt{\hat{u}_{C}^{2}+\beta/\rho}\ .
\end{equation}

\noindent The right eigenvectors of the Jacobian, written as columns
of a matrix, are,

\noindent \renewcommand{\arraystretch}{1.5}
\begin{equation}
R_{ev}=\left[\begin{array}{ccccc}
1 & 0 & 0 & 0 & 1\\
\frac{\rho\,\lambda_{1}\left(\hat{u}-\lambda_{4}\right)}{\lambda_{1}-\hat{u}} & 0 & 0 & 1 & \frac{\rho\,\lambda_{4}\,\left(\hat{u}-\lambda_{1}\right)}{\lambda_{4}-\hat{u}}\\
\frac{\rho\,\hat{v}\,\lambda_{1}}{\lambda_{1}-\hat{u}} & 1 & 0 & 0 & \frac{\rho\,\hat{v}\,\lambda_{4}}{\lambda_{4}-\hat{u}}\\
\frac{\rho\,\hat{w}\,\lambda_{1}}{\lambda_{1}-\hat{u}} & 0 & 1 & 0 & \frac{\rho\,\hat{w}\,\lambda_{4}}{\lambda_{4}-\hat{u}}\\
\frac{C\,\lambda_{1}}{\lambda_{1}-\hat{u}} & 0 & 0 & \frac{1}{\left(\rho_{1}-\rho_{2}\right)\,\hat{u}} & \frac{C\,\lambda_{4}}{\lambda_{4}-\hat{u}}
\end{array}\right]\ .
\end{equation}
\renewcommand{\arraystretch}{1}

\noindent There are five waves corresponding to each of the eigenvalues,
however, since three of the eigenvalues are repeating, it will result
in four distinct states separated by three waves as shown in Fig.~\ref{fig:HLLC-VOF-wave-structure}.
The solution states are denoted by $\mathbf{\hat{U}}_{L}$, $\mathbf{\hat{U}}_{*L}$,
$\mathbf{\hat{U}}_{*R}$, $\mathbf{\hat{U}}_{R}$ and the wave speeds
are denoted by $S_{L}$, $S_{*}$ and $S_{R}$.

\noindent 

\noindent This results in the following definition of flux depending
on the wave speeds,
\begin{equation}
\mathbf{\hat{F}}_{f}=\begin{cases}
\mathbf{\hat{F}}_{L} & S_{L}\geq0\\
\mathbf{\hat{F}}_{*L} & S_{L}\leq0\leq S_{*}\\
\mathbf{\hat{F}}_{*R} & S_{*}\leq0\leq S_{R}\\
\mathbf{\hat{F}}_{R} & S_{R}\leq0
\end{cases}\label{eq:Flux-Cases}
\end{equation}
where, 
\begin{equation}
\mathbf{\hat{F}}_{L}=\mathbf{F}\left(\mathbf{\hat{U}}_{L}\right)\ ;\quad\mathbf{\hat{F}}_{R}=\mathbf{F}\left(\mathbf{\hat{U}}_{R}\right)\ ;
\end{equation}
and using the Rankine-Hugoniot jump conditions across the left and
right wave,
\begin{equation}
\mathbf{\hat{F}}_{*L}=\mathbf{\hat{F}}_{L}+S_{L}\left(\mathbf{\hat{U}}_{*L}-\mathbf{\hat{U}}_{L}\right)\ ;\label{eq:FStarL}
\end{equation}
\begin{equation}
\mathbf{\hat{F}}_{*R}=\mathbf{\hat{F}}_{R}+S_{R}\left(\mathbf{\hat{U}}_{*R}-\mathbf{\hat{U}}_{R}\right)\ .\label{eq:FStarR}
\end{equation}
The intermediate states, $\mathbf{\hat{U}}_{*L}$, $\mathbf{\hat{U}}_{*R}$,
and the intermediate wave speed, $S_{*}$, can be calculated by using
the Rankine-Hugoniot jump conditions and imposing additional conditions
obtained from generalized Riemann invariant analysis. After a considerable
amount of mathematical rigor all the above unknowns can be estimated.
All the expressions necessary for calculation of convective flux are
concisely given below.
\begin{equation}
\mathbf{\hat{F}}_{L}=\mathbf{F}\left(\hat{\mathbf{U}}_{L}\right)=\left[\begin{array}{c}
\hat{u}\\
\rho\,\hat{u}^{2}+p\\
\rho\,\hat{u}\,\hat{v}\\
\rho\,\hat{u}\,\hat{w}\\
C\,\hat{u}
\end{array}\right]_{L};\qquad\mathbf{\hat{F}}_{R}=\mathbf{F}\left(\mathbf{\hat{U}}_{R}\right)=\left[\begin{array}{c}
\hat{u}\\
\rho\,\hat{u}^{2}+p\\
\rho\,\hat{u}\,\hat{v}\\
\rho\,\hat{u}\,\hat{w}\\
C\,\hat{u}
\end{array}\right]_{R}
\end{equation}
\begin{equation}
\boxed{\rho_{L}=\left(\rho_{1}-\rho_{2}\right)\left[C\right]_{L}+\rho_{2}}\qquad\boxed{\rho_{R}=\left(\rho_{1}-\rho_{2}\right)\left[C\right]_{R}+\rho_{2}}
\end{equation}
\begin{equation}
\hat{u}_{CL}=\left[\hat{u}\right]_{L}\,\frac{\left(\rho_{L}+\rho_{2}\right)}{2\,\rho_{L}}\qquad\hat{u}_{CR}=\left[\hat{u}\right]_{R}\,\frac{\left(\rho_{R}+\rho_{2}\right)}{2\,\rho_{R}}
\end{equation}
\begin{equation}
a_{L}=\sqrt{\hat{u}_{CL}^{2}+\beta/\rho_{L}};\qquad a_{R}=\sqrt{\hat{u}_{CR}^{2}+\beta/\rho_{R}}
\end{equation}
\begin{equation}
\boxed{S_{L}=\min\left(\hat{u}_{CL}-a_{L},\hat{u}_{CR}-a_{R}\right)}\qquad\boxed{S_{R}=\max\left(\hat{u}_{CL}+a_{L},\hat{u}_{CR}+a_{R}\right)}
\end{equation}
\begin{equation}
\boxed{\left[p/\beta\right]_{*L}=\left[p/\beta\right]_{*R}=\left[p/\beta\right]_{*}=\frac{\left[\hat{u}\right]_{L}-\left[\hat{u}\right]_{R}+S_{R}\left[p/\beta\right]_{R}-S_{L}\left[p/\beta\right]_{L}}{S_{R}-S_{L}}}
\end{equation}
\begin{equation}
\boxed{\hat{u}_{*L}=\hat{u}_{*R}=\hat{u}_{*}=S_{*}=\frac{S_{R}\left[\rho\,\hat{u}\right]_{R}-S_{L}\left[\rho\,\hat{u}\right]_{L}-\left[\rho\,\hat{u}^{2}+p\right]_{R}+\left[\rho\,\hat{u}^{2}+p\right]_{L}}{S_{R}\rho_{R}-S_{L}\rho_{L}-\left(\rho_{1}-\rho_{2}\right)\left(\left[C\,\hat{u}\right]_{R}-\left[C\,\hat{u}\right]_{L}\right)}}
\end{equation}
\begin{equation}
\boxed{\left[C\right]_{*L}=\frac{S_{L}\left[C\right]_{L}-\left[C\,\hat{u}\right]_{L}}{S_{L}-S_{*}}}\qquad\boxed{\left[C\right]_{*R}=\frac{S_{R}\left[C\right]_{R}-\left[C\,\hat{u}\right]_{R}}{S_{R}-S_{*}}}
\end{equation}
\begin{equation}
\boxed{\rho_{*L}=\left(\rho_{1}-\rho_{2}\right)\left[C\right]_{*L}+\rho_{2}}\qquad\boxed{\rho_{*R}=\left(\rho_{1}-\rho_{2}\right)\left[C\right]_{*R}+\rho_{2}}
\end{equation}
\begin{equation}
\boxed{\left[\rho\,\hat{u}\right]_{*L}=\rho_{*L}\hat{u}_{*}}\qquad\boxed{\left[\rho\,\hat{u}\right]_{*R}=\rho_{*R}\hat{u}_{*}}
\end{equation}
\begin{equation}
\boxed{\left[\rho\,\hat{v}\right]_{*L}=\frac{S_{L}\left[\rho\,\hat{v}\right]_{L}-\left[\rho\,\hat{u}\,\hat{v}\right]_{L}}{S_{L}-S_{*}}}\qquad\boxed{\left[\rho\,\hat{v}\right]_{*R}=\frac{S_{R}\left[\rho\,\hat{v}\right]_{R}-\left[\rho\,\hat{u}\,\hat{v}\right]_{R}}{S_{R}-S_{*}}}
\end{equation}
\begin{equation}
\boxed{\left[\rho\,\hat{w}\right]_{*L}=\frac{S_{L}\left[\rho\,\hat{w}\right]_{L}-\left[\rho\,\hat{u}\,\hat{w}\right]_{L}}{S_{L}-S_{*}}}\qquad\boxed{\left[\rho\,\hat{w}\right]_{*R}=\frac{S_{R}\left[\rho\,\hat{w}\right]_{R}-\left[\rho\,\hat{u}\,\hat{w}\right]_{R}}{S_{R}-S_{*}}}
\end{equation}
The terms in square brackets $\left[\ \right]$ indicate either conservative
variables or convective flux variables, which are stored in corresponding
vectors and can be used without additional computation. The intermediate
states are obtained by assembling the above terms as, 
\begin{equation}
\mathbf{U}_{*L}=\left[\begin{array}{c}
p/\beta\\
\rho\,\hat{u}\\
\rho\,\hat{v}\\
\rho\,\hat{w}\\
C
\end{array}\right]_{*L}\qquad\mathbf{U}_{*R}=\left[\begin{array}{c}
p/\beta\\
\rho\,\hat{u}\\
\rho\,\hat{v}\\
\rho\,\hat{w}\\
C
\end{array}\right]_{*R}\ .
\end{equation}
The convective flux in the rotated coordinate system can then be computed
using equations (\ref{eq:Flux-Cases}), (\ref{eq:FStarL}) and (\ref{eq:FStarR}).

Finally, to compute the required flux in the finite volume formulation
in the original coordinate system $\left(x,y,z\right)$, the inverse
transform can be applied,
\begin{equation}
\mathbf{F}\, n_{x}+\mathbf{G}\, n_{y}+\mathbf{H}\, n_{z}=T^{-1}\mathbf{\hat{F}}_{f}\ .\label{eq:transformed-flux}
\end{equation}
Since the transformation matrix, $T$, is orthonormal, its inverse
can be calculated by simply transposing the matrix,
\begin{equation}
T^{-1}=\left[\begin{array}{ccccc}
1 & 0 & 0 & 0 & 0\\
0 & n_{x} & t_{1x} & t_{2x} & 0\\
0 & n_{y} & t_{1y} & t_{2y} & 0\\
0 & n_{z} & t_{1z} & t_{2z} & 0\\
0 & 0 & 0 & 0 & 1
\end{array}\right]\ .
\end{equation}
The HLLC-VOF-M Riemann solver can be used for solving 2D and 3D problems,
using structured or unstructured mixed cells, without any modifications
in the formulation.

\subsection{\label{sub:interface-compression}Interface compression}

Approximate Riemann solvers inherently introduce some numerical dissipation
to ensure stability. This is acceptable in case of compressible fluid
flows, where Riemann solvers are extensively used, as the governing
equations itself aids compression of discontinuities, such as shocks,
to produce correct solutions. The passively advected volume fraction,
on the other hand, will keep spreading out the discontinuous volume
fraction at the interface, if not corrected artificially. To ensure
sharp interfaces, an interface compression term, $\nabla\cdot\mathbf{F}_{c}$,
is added to the governing equations (\ref{eq:AC_VOF-NS}). This technique
is borrowed from \cite{Rusche2002}, which is also implemented in
OpenFOAM \cite{Jasak2009}, and modified based on later advancements
in the technique \cite{Lee2015,Mehmani2018}. The compression term
can be written as, 
\[
\nabla\cdot\mathbf{F}_{c}=\left[0,0,0,0,\nabla\cdot\vec{V}_{c}\,\left(1-C\right)\, C\right]^{T}\ .
\]
The resulting VOF equation becomes,
\begin{equation}
\frac{\partial C}{\partial\tau}+\frac{\partial C}{\partial t}+\nabla\cdot\left(\vec{V}\, C+\vec{V}_{c}\,\left(1-C\right)\, C\right)=0\ ,
\end{equation}
where, $\vec{V}=\left(u,v,w\right)$ is the velocity field of the
fluids and $\vec{V}_{c}$ is the compressive velocity field. The term
$\left(1-C\right)\, C$, ensures that the interface compression is
active only near the interface.

The advection part of the flux, due to the term $\vec{V}\, C$, is
calculated by the HLLC-VOF-M Riemann solver, therefore only the compressive
part of the flux, due to the term $\vec{V}_{c}\,\left(1-C\right)\, C$,
needs to be computed. The remaining equation, without $\vec{V}\, C$,
can be viewed like an advection equation of $C$, with $\vec{V}_{c}\,\left(1-C\right)$
as the advection velocity field. The required compressive flux can
therefore be calculated using a simple scalar upwind solver using
the left and right states of volume fraction, $C_{L}$ and $C_{R}$,
\begin{equation}
F_{fc}=\begin{cases}
\lambda_{c}\, C_{L} & \lambda_{c}>0\ ;\\
\lambda_{c}\, C_{R} & \text{otherwise}\ .
\end{cases}\label{eq:flux-interface-compression}
\end{equation}
The wave speed is calculated as $\lambda_{c}=\left(1-C_{f}\right)\vec{V}_{c}\cdot\hat{n}$.
The volume fraction at the cell face is denoted as $C_{f}$. The compressive
velocity field is computed based on literature \cite{Rusche2002}
as,
\begin{equation}
\vec{V}_{c}=\zeta\,\eta\,\left|\vec{V}_{f}\cdot\hat{n}\right|\hat{n}_{i}\ ,
\end{equation}
where, $\zeta$ is the compressive coefficient which can be chosen
in the range of 0 to 1. In this work, a value of $\zeta=0.3$ is used.
$\eta$ is called the dynamic interface compression coefficient which
is given as $\eta=\sqrt{\left|\hat{n}\cdot\hat{n}_{i}\right|}$. The
unit vector normal to cell face is $\hat{n}$ and $\hat{n}_{i}$ is
the unit vector perpendicular to fluid interface, calculated as, 
\begin{equation}
\hat{n}_{i}=\begin{cases}
\nabla C_{f}/\left|\nabla C_{f}\right| & \text{if}\ \left|\nabla C_{f}\right|>10^{-6}\ ,\\
0 & \text{otherwise}\ ,
\end{cases}
\end{equation}
to avoid division by zero. The velocity at the cell face, $\vec{V}_{f}$;
the volume fraction at the cell face, $C_{f}$; and the gradient of
volume fraction at the cell face, $\nabla C_{f}$, are calculated
as a part of viscous flux calculation (explained in section \ref{sub:viscous-flux})
and are reused for interface compression. The value obtained by equation
(\ref{eq:flux-interface-compression}) is simply added to the fifth
component of flux obtained by the HLLC-VOF-M Riemann solver (or HLL
Riemann solver, when used for comparison) to avoid smearing of the
interface.

\subsection{\label{sub:viscous-flux}Calculation of viscous flux}

\begin{figure}[h]
\begin{centering}
\includegraphics[width=0.5\textwidth,draft=false]{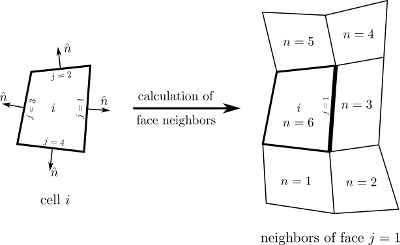}
\par\end{centering}

\caption{\label{fig:face-neighbours}An example of face neighbors for numerical
calculation of gradients at the cell face $j=1$}
\end{figure}

In the finite volume discretization, the viscous flux needs to be
calculated at the cell faces. Hence, the velocity gradients, $\nabla V$,
and the dynamic coefficient of viscosity, $\mu$, need to be computed
at the cell faces. The velocity gradients can be written as,
\begin{equation}
\nabla V=\left[\begin{array}{ccc}
\partial u/\partial x & \partial u/\partial y & \partial u/\partial z\\
\partial v/\partial x & \partial v/\partial y & \partial v/\partial z\\
\partial w/\partial x & \partial w/\partial y & \partial w/\partial z
\end{array}\right]\ .
\end{equation}
For numerical computation of derivatives at the face, cell values
in the vicinity of the face are needed. Consider the cell $i$ shown
in Fig.~\ref{fig:face-neighbours}. The set of neighbors of a face,
say $j=1$, is constructed by collecting the neighbors of all the
nodes of the face. In the example shown in Fig.~\ref{fig:face-neighbours},
this will result in a set of 6 neighbors i.e. cells $n=1\dots5$ and
cell $i$. Here a quadrilateral cell is considered, however the same
strategy can be followed for all types of cells, in two- and three-dimensions.

\noindent 

In a finite volume formulation, as we know the cell averaged values
of the conservative variables, the required derivatives of velocity
are calculated in two steps. The first step is to compute the gradients
of the conservative variables at the face; and the second step is
to convert the gradients of conservative variables to gradients of
velocity by using the Jacobian of transformation. The gradients of
the conservative variables are calculated by using a truncated Taylor
series. Consider a variable, $\xi$, to be a component of the conservative
variable vector $\mathbf{U}$. A truncated Taylor series to obtain
a linear approximation of $\xi$ can be written as,
\begin{equation}
\xi_{n}=\xi_{j}+\left.\frac{\partial\xi}{\partial x}\right|_{j}\Delta x_{n}+\left.\frac{\partial\xi}{\partial y}\right|_{j}\Delta y_{n}+\left.\frac{\partial\xi}{\partial z}\right|_{j}\Delta z_{n}\ ,
\end{equation}
where, the derivatives are evaluated at the centroid of face $j$.
$\left(\Delta x_{n},\Delta y_{n},\Delta z_{n}\right)$ is the vector
from the centroid of face $j$ to the centroid of neighbor cell $n$.
Writing the above equation for all the neighbors, $n=1,2\dots N$,
of a face will result in the following system:
\begin{equation}
\underbrace{\left[\begin{array}{cccc}
1 & \Delta x_{1} & \Delta y_{1} & \Delta z_{1}\\
1 & \Delta x_{2} & \Delta y_{2} & \Delta z_{2}\\
\vdots & \vdots & \vdots & \vdots\\
1 & \Delta x_{N} & \Delta y_{N} & \Delta z_{N}
\end{array}\right]}_{S}\cdot\underbrace{\left[\begin{array}{c}
\xi\\
\partial\xi/\partial x\\
\partial\xi/\partial y\\
\partial\xi/\partial z
\end{array}\right]_{j}}_{\xi_{j}^{\prime}}=\underbrace{\left[\begin{array}{c}
\xi_{1}\\
\xi_{2}\\
\vdots\\
\xi_{N}
\end{array}\right]}_{\xi_{n}}\ .
\end{equation}
The equations are weighted using inverse-distance based weights, $w_{n}=1/\left\Vert \Delta r_{n}\right\Vert $,
where $\left\Vert \Delta r_{n}\right\Vert =\sqrt{\Delta x_{n}^{2}+\Delta y_{n}^{2}+\Delta z_{n}^{2}}$.
The system of equations with weights can be written as,
\begin{equation}
W\, S\,\xi_{j}^{\prime}=W\,\xi_{n}\ ,\label{eq:ls-face-gradients-conservative-variables}
\end{equation}
with $W=\text{diag}\left(w_{1},w_{2}\dots w_{N}\right)$. This results
in a tall system of equations with four unknowns, $\xi$, $\partial\xi/\partial x$,
$\partial\xi/\partial y$ and $\partial\xi/\partial z$ at the centroid
of face $j$. The system is solved by using singular value decomposition
(SVD) method using an open source mathematical library \cite{ApacheCommonsMath}. 

The above procedure can be followed for each component of the conservative
variable vector, $\mathbf{U}$, to obtain its value and gradient at
the cell face. The gradients of conservative variables at a face $j$,
can be packed together as,
\begin{equation}
\nabla\mathbf{U}_{j}=\left[\begin{array}{ccc}
\partial U_{1}/\partial x & \partial U_{1}/\partial y & \partial U_{1}/\partial z\\
\partial U_{2}/\partial x & \partial U_{2}/\partial y & \partial U_{2}/\partial z\\
\partial U_{3}/\partial x & \partial U_{3}/\partial y & \partial U_{3}/\partial z\\
\partial U_{4}/\partial x & \partial U_{4}/\partial y & \partial U_{4}/\partial z\\
\partial U_{5}/\partial x & \partial U_{5}/\partial y & \partial U_{5}/\partial z
\end{array}\right]\ ,
\end{equation}
where, $U_{1}=p/\beta$, $U_{2}=\rho\, u$, $U_{3}=\rho\, v$, $U_{4}=\rho\, w$,
$U_{5}=C$.

Once the gradients of the conservative variables at the cell face
are computed, the second step is to calculate the velocity gradients
by using the transformation,
\begin{equation}
\nabla V_{j}=J\,\nabla\mathbf{U}_{j}\ ,
\end{equation}
where $J$ is the Jacobian, obtained as,\renewcommand{\arraystretch}{1.5}
\begin{equation}
J=\frac{\partial V}{\partial\mathbf{U}}=\left[\begin{array}{ccccc}
0 & 1/\rho_{j} & 0 & 0 & \frac{\left(\rho_{2}-\rho_{1}\right)u_{j}}{\rho_{j}}\\
0 & 0 & 1/\rho_{j} & 0 & \frac{\left(\rho_{2}-\rho_{1}\right)v_{j}}{\rho_{j}}\\
0 & 0 & 0 & 1/\rho_{j} & \frac{\left(\rho_{2}-\rho_{1}\right)w_{j}}{\rho_{j}}
\end{array}\right]\ .
\end{equation}
\renewcommand{\arraystretch}{1}As a part of the solution of system
(\ref{eq:ls-face-gradients-conservative-variables}), the interpolated
values of conservative variables, $\left(\rho u\right)_{j}$, $\left(\rho v\right)_{j}$,
$\left(\rho w\right)_{j}$ and $C_{j}$, are also obtained at the
cell face. The density, $\rho_{j}$, at the cell face can be calculated
using equation (\ref{eq:density_function}). The velocity components
at the cell face can be calculated as,
\begin{equation}
u_{j}=\left(\rho u\right)_{j}/\rho_{j},\quad v_{j}=\left(\rho v\right)_{j}/\rho_{j},\quad w_{j}=\left(\rho w\right)_{j}/\rho_{j}\ ,
\end{equation}
which can be used for calculation of the Jacobian, $J$. The value
of the coefficient of dynamic viscosity, $\mu_{j}$, at the cell face
can be calculated using equation (\ref{eq:dyn_visc_function}). The
viscous flux can be calculated using the obtained gradients of velocity
and coefficient of dynamic viscosity.

\subsection{\label{sub:surface-tension}Calculation of surface tension flux}

Surface tension is a force which acts only at the interface between
two fluids, which is a result of intermolecular forces. This force
acts in such a way so as to minimize the surface area of the interface.
The magnitude of the resultant force is proportional to surface tension
coefficient, $\sigma$, corresponding to the fluids. The value of
$\sigma$ can be obtained experimentally for different fluid combinations
from literature \cite{Adamson1997}. This force is commonly modeled
in numerical simulations, using interface capturing methods, by the
continuum surface force (CSF) \cite{Brackbill1992} or the continuum
surface stress (CSS) \cite{Lafaurie1994} models. In this work, the
CSS method is used for modeling surface tension as it does not require
explicit calculation of interface curvature. Also, based on numerical
experiments it was observed by \cite{Lafaurie1994}, that the CSS
model does not require smoothening of VOF function. Hence, it is very
straightforward and quite natural to combine the CSS model within
the VOF framework in a conservative finite volume formulation. The
CSS model describes the surface tension force in a momentum equation
as,
\begin{equation}
\mathbf{F_{s}}=\sigma\,\left(\left|\nabla C\right|\ I-\frac{\nabla C\otimes\nabla C}{\left|\nabla C\right|}\right)\ ,
\end{equation}
where, $\sigma$ is the surface tension coefficient, $\nabla C$ is
the gradient of volume fraction, $\left|\nabla C\right|$ is the magnitude
of the gradient vector and $I$ is a $3\times3$ identity matrix.
The surface tension force in the governing equations (\ref{eq:AC_VOF-NS})
therefore becomes,
\[
\nabla\cdot\mathbf{T}=\left[\begin{array}{c}
0\\
\nabla\cdot\mathbf{F_{s}}\\
0
\end{array}\right]\ .
\]
The surface tension force is set to zero away from the interface,
explicitly, when the value of $\left|\nabla C\right|<10^{-6}$ to
avoid division by zero. The tensor $\nabla C\otimes\nabla C$ is defined
as,\renewcommand{\arraystretch}{1.8}
\begin{equation}
\nabla C\otimes\nabla C=\nabla C^{T}\nabla C=\left[\begin{array}{ccc}
\left(\frac{\partial C}{\partial x}\right)^{2} & \frac{\partial C}{\partial x}\frac{\partial C}{\partial y} & \frac{\partial C}{\partial x}\frac{\partial C}{\partial z}\\
\frac{\partial C}{\partial x}\frac{\partial C}{\partial y} & \left(\frac{\partial C}{\partial y}\right)^{2} & \frac{\partial C}{\partial y}\frac{\partial C}{\partial z}\\
\frac{\partial C}{\partial x}\frac{\partial C}{\partial z} & \frac{\partial C}{\partial y}\frac{\partial C}{\partial z} & \left(\frac{\partial C}{\partial z}\right)^{2}
\end{array}\right]\ .
\end{equation}
\renewcommand{\arraystretch}{1}

For incorporating the CSS model into finite volume formulation, the
value of $\mathbf{F_{s}}$ has to be computed at the cell face. The
gradients at the face,$\nabla\mathbf{U}_{j}$, are computed during
calculation of viscous flux, which is described in sub-section~\ref{sub:viscous-flux}.
The gradient of volume fraction for a face $j$ is calculated as,
\begin{equation}
\nabla C_{j}^{T}=J\,\nabla\mathbf{U}_{j}\ ,
\end{equation}
where $J$ can be obtained to be,
\begin{equation}
J=\frac{\partial C}{\partial\mathbf{U}}=\left[\begin{array}{ccccc}
0 & 0 & 0 & 0 & 1\end{array}\right]\ .
\end{equation}

\subsection{\label{sub:source-term}Treatment of source term}

The source term in the finite volume formulation is given by $\Omega_{P}\,\overline{\mathbf{S}}$,
with $\mathbf{\overline{S}}=\left[0,\overline{\rho}\, g_{x},\overline{\rho}\, g_{y},\overline{\rho}\, g_{z},0\right]^{T}$
and $\Omega_{P}$ is the volume of the cell. The average value of
the density, $\overline{\rho}$, can be computed simply by using equation
(\ref{eq:density_function}) and substituting the average value of
volume fraction, $\overline{C}$, which is available in the finite
volume formulation.

\subsection{Initial conditions}

An accurate calculation of the initial cell-averaged values of conservative
variables is important in two-phase flow problems, as the solution
is sensitive to initial position of the fluid interface. To obtain
the initial cell-averaged values, an integration of the initial distribution
needs to be computed over each cell. A numerical integration is carried
out, for each cell, by subdividing the cell into smaller parts and
then performing a Riemann sum. In case of quadrilateral (and hexahedral)
cells, the cell is divided by using transfinite interpolation with
20 divisions (4 divisions for hexahedral) along each computational
coordinate direction. This results in $20\times20$ sub-quadrilaterals
($4\times4\times4$ sub-hexahedrals) per cell. In case of triangular
cells, the sub-triangles are generated by using 5 levels of recursive
sub-division with each level diving a triangle into 4 equal sub-triangles.
This results in $1024$ sub-triangles per cell.

\subsection{Boundary conditions}

The boundary conditions are implemented using ghost cells (also known
as dummy cells). The ghost cells are additional finite volume cells
added outside the computational domain so that gradients at the boundary
can be specified or computed. There are three types of boundary conditions
used in this work: the stationary slip wall (also called inviscid
wall), stationary no-slip wall (also called viscous wall) and symmetry
boundary condition. In all these boundary conditions the normal component
of velocity is zero, therefore the convective flux is given by $\left[0,p\, n_{x},p\, n_{y},p\, n_{z},0\right]^{T}$.
The calculation of viscous and the surface tension flux at the boundary
depend on the gradients of variables at the boundary face, therefore
the ghost cells have to be set appropriately. In case of slip wall
boundary, the inside velocity vector is mirrored about the boundary
plane, while the pressure and volume fraction are extrapolated using
corresponding values and gradients from the inside cell. In case of
no-slip boundary, the negative of velocity vector is copied to the
ghost cell, while the pressure and volume fraction are extrapolated.
In case of symmetry, the inside velocity vector is mirrored about
the boundary plane, while the pressure and volume fraction are copied
as it is to the ghost cell.

\section{\label{sec:The-Algorithm}The Algorithm}

\begin{figure}[h]
\noindent \begin{centering}
\includegraphics[width=0.4\textwidth,draft=false]{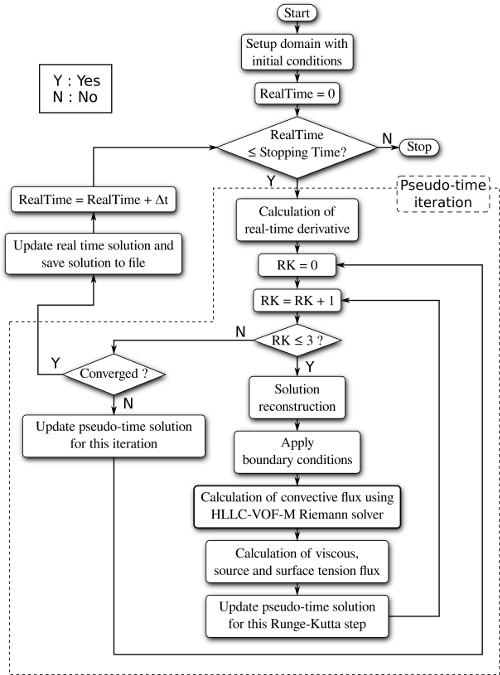}
\par\end{centering}

\caption{\label{fig:flowchart-algorithm}Flowchart of the complete algorithm}
\end{figure}

The Navier-Stokes equations are written using artificial compressibility
formulation and combined with the VOF equation to simulate transient
two-phase flows. A dual-time stepping algorithm is used to obtain
a time-accurate solution of the flow field. The dual-time stepping
method allows for implicit treatment of real-time variables, and therefore
the real-time step size is restricted only by the physics of the flow.
In the dual time stepping approach, the outer loop advances the real-time
variables, $\mathbf{W}$, in real-time, $t$. The inner loop iterates
in pseudo-time, $\tau$, and continues until the solution converges
to a time accurate solution, as the residual drops below acceptable
limits (which is chosen as $10^{-3}$ in this work). During the pseudo-time
iterations, as the solution evolves, the latest solution is used for
evaluating the total residual. The total residual is computed by summing
up the convective, interface compression, viscous, surface tension
and the source residual. The convective part of the residual is computed
using the HLLC-VOF-M Riemann solver, introduced in this paper. A flowchart
of the complete algorithm is displayed in Fig.~\ref{fig:flowchart-algorithm}.

\noindent 

\section{\label{sec:Results-and-discussion}Results and discussion}

To test the efficacy and practical applicability of the HLLC-VOF-M
convective flux solver, several two-dimensional problems are solved
using structured and unstructured mesh. Also, a three-dimensional
problem involving complex merging of interfaces is solved to demonstrate
the extensibility and robustness of the solver. The results are compared
with theoretical, experimental and numerical results available in
literature. In all the problems the artificial compressibility parameter,
$\beta$, is chosen to be $1000$. The velocity field of the flow
is displayed in some figures, using scaled arrows. The scaling of
these arrows is based on the largest velocity magnitude in the complete
domain at the time instant. The fluid interface is displayed by plotting
two levels at 0.3 and 0.7, in all the two-dimensional contour plots.
In three-dimensions, a contour level of 0.5 is used to display the
fluid interface.

\subsection{Droplet splash}

\begin{figure}[h]
\begin{centering}
\begin{tabular}{>{\centering}p{0.1\textwidth}>{\centering}p{0.1\textwidth}>{\centering}p{0.1\textwidth}>{\centering}p{0.1\textwidth}>{\centering}p{0.1\textwidth}>{\centering}p{0.1\textwidth}>{\centering}p{0.1\textwidth}>{\centering}p{0.1\textwidth}}
{\footnotesize{}\includegraphics[width=0.1\textwidth,draft=false]{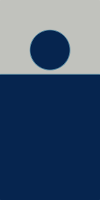}}{\footnotesize \par}

{\footnotesize{}$t=0.0\,\text{s}$}\\
 & {\footnotesize{}\includegraphics[width=0.1\textwidth,draft=false]{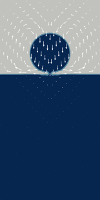}}{\footnotesize \par}

{\footnotesize{}$t=0.00677\,\text{s}$}\\
 & {\footnotesize{}\includegraphics[width=0.1\textwidth,draft=false]{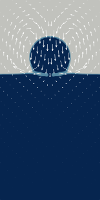}}{\footnotesize \par}

{\footnotesize{}$t=0.00980\,\text{s}$}\\
 & {\footnotesize{}\includegraphics[width=0.1\textwidth,draft=false]{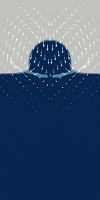}}{\footnotesize \par}

{\footnotesize{}$t=0.01220\,\text{s}$}\\
 & {\footnotesize{}\includegraphics[width=0.1\textwidth,draft=false]{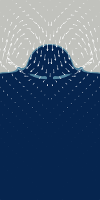}}{\footnotesize \par}

{\footnotesize{}$t=0.01485\,\text{s}$} & {\footnotesize{}\includegraphics[width=0.1\textwidth,draft=false]{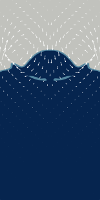}}{\footnotesize \par}

{\footnotesize{}$t=0.01781\,\text{s}$} & {\footnotesize{}\includegraphics[width=0.1\textwidth,draft=false]{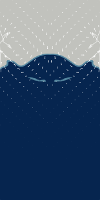}}{\footnotesize \par}

{\footnotesize{}$t=0.01995\,\text{s}$} & {\footnotesize{}\includegraphics[width=0.1\textwidth,draft=false]{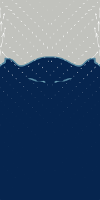}}{\footnotesize \par}

{\footnotesize{}$t=0.02146\,\text{s}$}\tabularnewline
\end{tabular}
\par\end{centering}

\caption{\label{fig:snapshots-drop-splash-hllcvof2}Snapshots of the drop splash
problem at various time levels, using HLLC-VOF-M solver.}
\end{figure}

\begin{figure}[h]
\begin{centering}
\begin{tabular}{>{\centering}p{0.1\textwidth}>{\centering}p{0.1\textwidth}>{\centering}p{0.1\textwidth}>{\centering}p{0.1\textwidth}>{\centering}p{0.1\textwidth}>{\centering}p{0.1\textwidth}>{\centering}p{0.1\textwidth}>{\centering}p{0.1\textwidth}}
{\footnotesize{}\includegraphics[width=0.1\textwidth]{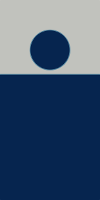}}{\footnotesize \par}

{\footnotesize{}$t=0.0\,\text{s}$}\\
 & {\footnotesize{}\includegraphics[width=0.1\textwidth]{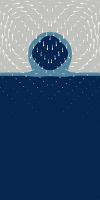}}{\footnotesize \par}

{\footnotesize{}$t=0.00677\,\text{s}$}\\
 & {\footnotesize{}\includegraphics[width=0.1\textwidth]{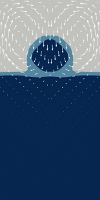}}{\footnotesize \par}

{\footnotesize{}$t=0.00980\,\text{s}$}\\
 & {\footnotesize{}\includegraphics[width=0.1\textwidth]{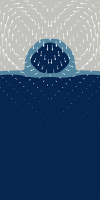}}{\footnotesize \par}

{\footnotesize{}$t=0.01220\,\text{s}$}\\
 & {\footnotesize{}\includegraphics[width=0.1\textwidth]{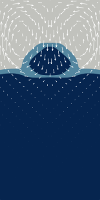}}{\footnotesize \par}

{\footnotesize{}$t=0.01485\,\text{s}$} & {\footnotesize{}\includegraphics[width=0.1\textwidth]{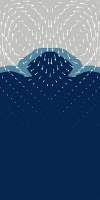}}{\footnotesize \par}

{\footnotesize{}$t=0.01781\,\text{s}$} & {\footnotesize{}\includegraphics[width=0.1\textwidth]{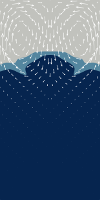}}{\footnotesize \par}

{\footnotesize{}$t=0.01995\,\text{s}$} & {\footnotesize{}\includegraphics[width=0.1\textwidth]{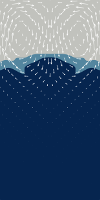}}{\footnotesize \par}

{\footnotesize{}$t=0.02146\,\text{s}$}\tabularnewline
\end{tabular}
\par\end{centering}

\caption{\label{fig:snapshots-drop-splash-hll}Snapshots of the drop splash
problem at various time levels, using HLL solver.}
\end{figure}

In this problem, a two-dimensional circular water droplet is placed
over bulk water body separated by air. Numerical results of this problem
are available in literature \cite{Puckett1997}, where high order
pressure-based, projection method is used. The effect of surface tension
is not considered in the said paper, hence in this problem the surface
tension coefficient is set to zero. The properties of the two fluids
considered are water ($\rho_{1}=998\,\text{kg/m}^{3},\mu_{1}=1.002\times10^{-3}\,\text{Pa-s}$)
and air ($\rho_{2}=1.2\,\text{kg/m}^{3},\mu_{2}=1.825\times10^{-5}\,\text{Pa-s}$).
The acceleration due to gravity is $g=\left(0,-9.81,0\right)\,\text{m/s}^{2}$.
The domain is of size $0.007\,\text{m}\times0.014\,\text{m}$ with
water filled up to $0.0088\,\text{m}$. The water droplet is placed
above the water surface with its center at $\left(0.0035\,\text{m},\ 0.0105\,\text{m}\right)$
and having a diameter of $2.8\times10^{-3}\,\text{m}$. The domain
is discretized using a Cartesian mesh of size $80\times160$. The
initial velocity and pressure in the entire domain is zero. The problem
is solved up to a time of $t=0.025\,\text{s}$ using steps of $\Delta t=10^{-5}\,\text{s}$.
The initial distribution of volume fraction along with the snapshots
of the solution at various other time steps (same as \cite{Puckett1997})
using HLLC-VOF-M and HLL are shown in Fig.~\ref{fig:snapshots-drop-splash-hllcvof2}
and Fig.~\ref{fig:snapshots-drop-splash-hll} respectively. It can
be seen that the results obtained by the non-contact capturing HLL
solver produces excessive numerical dissipation and the tiny air bubbles
trapped inside the water are not captured by the solution. On the
other hand, the results obtained using HLLC-VOF-M solver agree very
well with the results from the literature \cite{Puckett1997}.

\noindent 

\noindent 

\subsection{Three-dimensional non-axisymmetric merging of bubbles}

In this problem, two spherical bubbles are placed slightly offset
to one another in a tank of size $1\,\text{m}\times2\mbox{\,\text{m}}\times1\,\text{m}$.
The diameter of the bubbles is $0.4\,\text{m}$. The lower bubble
is centered at $\left(0.6\,\text{m},0.25\,\text{m},0.6\,\text{m}\right)$;
and the upper bubble is centered at $\left(0.4\,\text{m},0.65\,\text{m},0.4\,\text{m}\right)$,
measured from the bottom corner. Due to buoyancy the bubbles start
moving up, and merge together as time evolves. A similar problem is
solved numerically, in literature \cite{Unverdi1992}, using the advancing
front method.

\begin{figure}[h]
\begin{centering}
\begin{tabular}{>{\centering}p{0.1\textwidth}>{\centering}p{0.1\textwidth}>{\centering}p{0.1\textwidth}>{\centering}p{0.1\textwidth}>{\centering}p{0.1\textwidth}>{\centering}p{0.1\textwidth}>{\centering}p{0.1\textwidth}>{\centering}p{0.1\textwidth}}
{\footnotesize{}\includegraphics[width=0.1\textwidth]{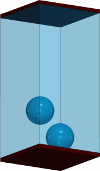}}{\footnotesize \par}

{\footnotesize{}$t=0.0\,\text{s}$}\\
 & {\footnotesize{}\includegraphics[width=0.1\textwidth]{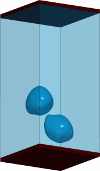}}{\footnotesize \par}

{\footnotesize{}$t=0.2\,\text{s}$}\\
 & {\footnotesize{}\includegraphics[width=0.1\textwidth]{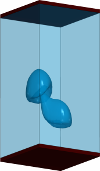}}{\footnotesize \par}

{\footnotesize{}$t=0.4\,\text{s}$}\\
 & {\footnotesize{}\includegraphics[width=0.1\textwidth]{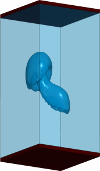}}{\footnotesize \par}

{\footnotesize{}$t=0.6\,\text{s}$}\\
 & {\footnotesize{}\includegraphics[width=0.1\textwidth]{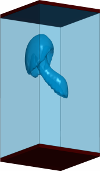}}{\footnotesize \par}

{\footnotesize{}$t=0.8\,\text{s}$} & {\footnotesize{}\includegraphics[width=0.1\textwidth]{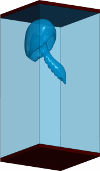}}{\footnotesize \par}

{\footnotesize{}$t=1.0\,\text{s}$} & {\footnotesize{}\includegraphics[width=0.1\textwidth]{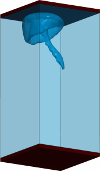}}{\footnotesize \par}

{\footnotesize{}$t=1.2\,\text{s}$} & {\footnotesize{}\includegraphics[width=0.1\textwidth]{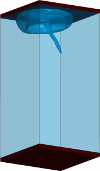}}{\footnotesize \par}

{\footnotesize{}$t=1.4\,\text{s}$}\tabularnewline
\end{tabular}
\par\end{centering}

\caption{\label{fig:snapshots-two-bubble-rise}Snapshots of two non-axisymmetric
rising bubbles at different times, computed using HLLC-VOF-M Riemann
solver. Mesh = $32\times64\times32$ cells, $Eo=50$, $M=1$, $\rho_{1}/\rho_{2}=20$
and $\mu_{1}/\mu_{2}=26$.}
\end{figure}

The bubbles and the outer fluid can be characterized by using the
Eötvös number $Eo=\rho_{1}\, g\, d_{e}^{2}/\sigma$, the Morton number
$M=g\,\mu_{1}^{4}/\left(\rho_{1}\,\sigma\right)$, and the ratio of
fluid properties $\rho_{1}/\rho_{2}$ and $\mu_{1}/\mu_{2}$. Subscript
1 is used for heavier outer fluid and subscript 2 is used for the
fluid contained inside the bubbles. The acceleration due to gravity
in the downward direction is chosen as, $g=1\,\text{m/s}^{2}$; the
density of the outer fluid is chosen as, $\rho_{1}=1\,\text{kg/m}^{3}$;
and the effective diameter of the bubbles is $d_{e}=0.4\,\text{m}$.
The simulation is performed for $Eo=50$, $M=1$, $\rho_{1}/\rho_{2}=20$
and $\mu_{1}/\mu_{2}=26$. The remaining fluid properties can be easily
evaluated using the above equations. The domain is discretized using
a mesh of $32\times64\times32$ cells. The pressure and velocity are
initialized to zero in the entire domain. All the boundaries of the
domain are set as slip-walls. The simulation is carried out for a
time duration of $t=5\,\text{s}$ with a time step of $\Delta t=0.001\,\text{s}$.
Various snapshots of the solution at different times of evolution
are shown in Fig.~\ref{fig:snapshots-two-bubble-rise}. The solutions
from literature \cite{Unverdi1992} are visually similar to the solutions
displayed in Fig.~\ref{fig:snapshots-two-bubble-rise}.

\noindent 

\section{\label{sec:Conclusion}Conclusion}

An improved HLLC-type Riemann solver is developed for three-dimensional,
incompressible two-phase fluid flow. The flow is modeled using the
artificial compressibility formulation and volume of fluid method.
The convective flux is derived for the system of governing equations
using the Rankine-Hugoniot jump conditions. The generalized Riemann
invariant analysis is used to apply the necessary additional constraints.
The convective flux computation using an upwind type solver inherently
introduces numerical dissipation, which causes smearing of the fluid
interface. Therefore, an interface compression mechanism is adopted
to overcome the unphysical dissipation effectively. The viscous flux
is computed using weighted least squares, the surface tension is computed
using the CSS model, and the source term is computed using the cell
average values. The solver is generic and can be used to solve incompressible
two-phase flow problems using structured and unstructured meshes in
two- and three-dimensions, without any modifications. The new Riemann
solver is found to be robust and performs well in all the selected
test problems with a single chosen value of artificial compressibility
parameter, $\beta$, unlike our previous formulation \cite{Bhat2019},
where the value of $\beta$ was determined based on trial-and-error.
Various two- and three-dimensional test problems are solved on different
mesh types to demonstrate the efficacy and robustness of the new solver.

\bibliographystyle{ieeetr}
\bibliography{/home/sourabh/Research/Annual_Progress_Seminar/References}

\end{document}